\documentstyle[epsf]{elsart}
\begin{document}
\begin{frontmatter}
\title{QED in strong fields}
\thanks{Invited talk at the workshop on
`Channeling and Other Coherent Crystal Effects at Relativistic Energies',
Aarhus, Denmark, July 10-14, 1995}
\author{A. Sch\"afer}
\address{Institut f\"ur Theoretische Physik, University of Frankfurt,
D-60054 Frankfurt}

\begin{abstract}
We discuss the status of atomic physics in strong field.
We focus on the problem of electron-positron lines observed in heavy-ion
collisions and on
QED effects, calculated in strong Coulomb fields, especially
Delbr\"uck-scattering.
We discuss the similarities and differences between these effects and
channeling respectively beamstrahlung.
We investigated the prospects for photon-channeling, calculated
channeling from first principles on the basis of the Dirac equation,
and determined the rate for electron-positron pair production in the
collision of two high-energy particle pulses.
\end{abstract}
\end{frontmatter}

In this contribution I would like to put the channeling phenomenon in a larger
context, namely the physics of strong electromagnetic fields, a subject
extensively investigated in Frankfurt since many years. The common technical
feature of the many different effects investigated under this label is that
the interaction of (usually) electrons with the electromagnetic fields of
the problem cannot be treated in perturbation theory, but has to be taken
into account exactly \cite{Gr1}.\\
To illustrate the diversity of this field I would like to address very
shortly three
topic, namely the status of the electron-positron coincidences observed
at GSI for quite a number of years, some problems of QED in heavy atoms, and
our contribution to the understanding of channeling and beamstrahlung.\\

\section{Electron-positron production in heavy-ion collisions}
In the collision of two heavy ions one can generate for a very short time
an atomic system with an effective charge which is the sum of the
charges of the
colliding nuclei. This is possible because electrons with a wavelength
of 1/$m_{\rm e}\approx 400$ fm cannot resolve structures far below this scale.
For the short time such a collision takes an electron becomes very strongly
bound.
Its binding energy can actually exceed twice its rest mass. This happens
if the combined atomic charge is larger than roughly 173, depending on
the amount of shielding by the other electrons. Fig.1 gives an impression of
the
relevant atomic processes in such an `overcritical' heavy ion collisons.
\begin{figure}[htb]
\epsffile{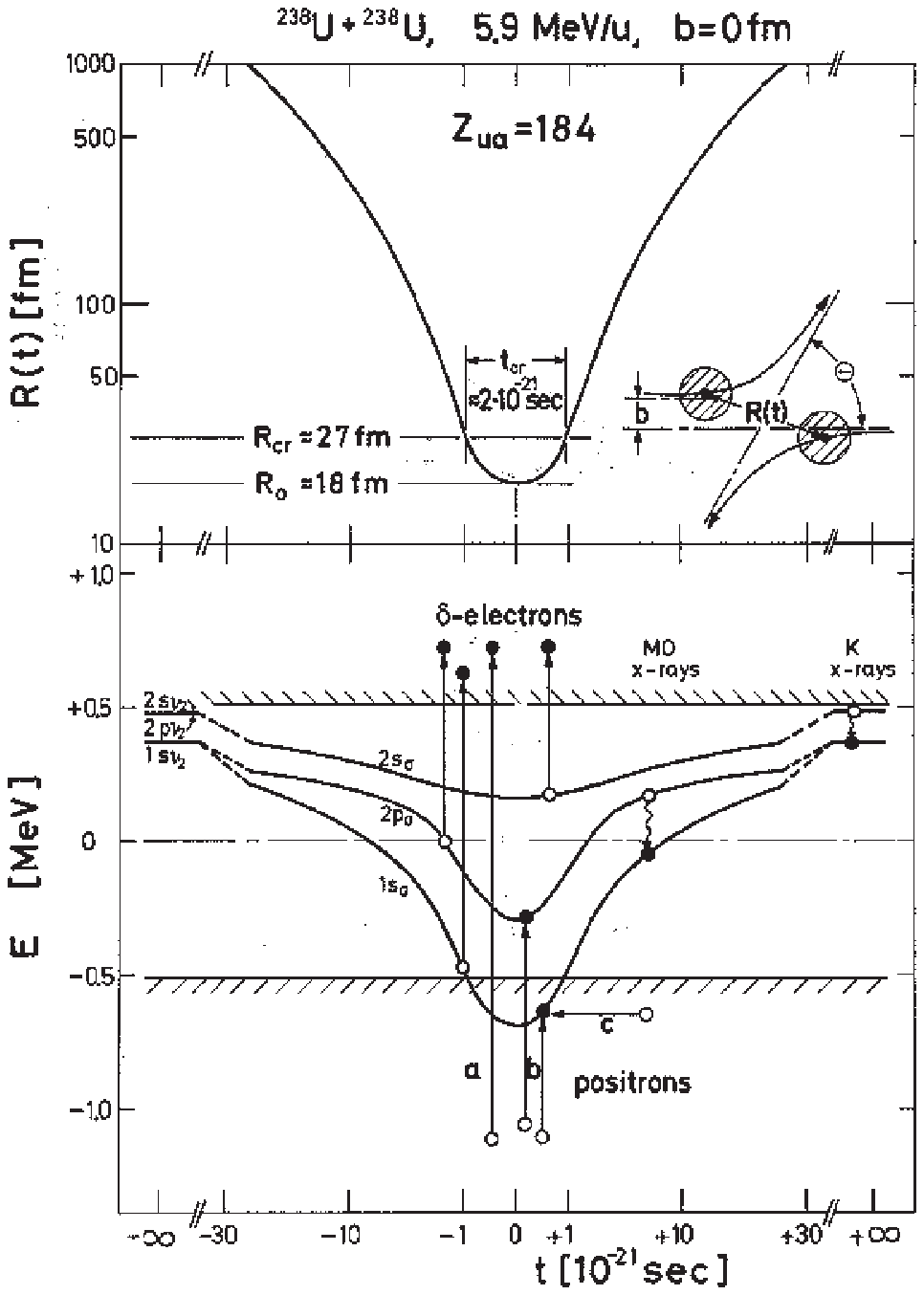}
\caption{The basic processes taking place during an
overcritical heavy ion collision}
\end{figure}
These
processes have been extensively studied, both theoretically and
experimentally. With the exception of  $\delta$-electron production
in U+Pd collisions at 6.1 MeV/u the theoretical and experimental results
have been found to
agree perfectly  within their uncertainties, which are typically 10 - 20
percent.\\
The technical treatment of strong field problems is basically always the same.
Because the Coulomb-potential is so strong it cannot be treated as
perturbation
but has to be included into the fundamental hamiltonian. For
the special case of
two colliding heavy ions this reads
\begin{equation}
H_0(\vec R(t))= \vec\alpha \cdot (\vec p -e\vec A) + \beta m +
e V_{\rm Cb}(\vec R,\vec r) + e V_{\rm e}(\vec R,\vec r)
\end{equation}
with
\begin{equation}
V_{\rm Cb}(\vec r)=- \int_{V_p} d^3r_p~ {\rho_p(\vec r_p)\over
|\vec r -\vec r_p|}
- \int_{V_t} d^3r_t~ {\rho_t(\vec r_t)\over |\vec r -\vec r_t|}.
\end{equation}
Here $\rho$ is the nuclear charge density and the indices $p$ and $t$ stand
for projectile and target. It is also
practical to include the dominant part of the electron-eclectron interaction,
which describes shielding of the ion charges, in the potential. Thus
$ V_{\rm e}$ can, e.g., be chosen as
\begin{eqnarray}
V_{\rm e}&= \sum_k \int d^3r'~ \Phi^{\dagger}_k(\vec r\, ')
{e\over |\vec r-\vec r\, '|} \Phi_k(\vec r\, ')\left|_{\rm monopole} \right.
\nonumber \\
 &- {2e\over 3} \left( {81\over 8 \pi} \right)^{1/3} \sum_k
 \Phi^{\dagger}_k(\vec r)
 \Phi_k(\vec r) \left|_{\rm monopole} \right.
\end{eqnarray}
implying a Slater approximation for the exchange term. In writing down this
hamiltonian we have made the monopole approximation, i.e. higher multipoles
of the
Coulomb-field are neglected. One can go beyond this approximation, but it was
found that the contributions from higher multipoles are negligable.
$H_0$ is used to calculate the time-independent eigenstates of the two-ion
system. Time dependence enters through the varying
distance vector between the two ions
$\vec R(t)$.
The time-dependent Dirac-equation can be converted to a system of coupled
differential equations
by expanding the wave functions according to
\begin{equation}
\Psi_i(\vec R(t))=\sum_j a_{ij}(t) \Phi_j(\vec R)\exp(-i \chi_j(t))
\end{equation}
\begin{equation}
\chi_j(t)=\int^tdt' \langle \Phi_j|H_0(\vec R(t'))|\Phi_j\rangle
\end{equation}
\begin{equation}
\stackrel{\bf \cdot}{a}_{ij}(t)= -\sum_{k\neq j}a_{ik}\langle \Phi_j|
{\partial\over \partial t}+iH_0(t)|\Phi_k\rangle
\exp(-i (\chi_j(t)-\chi_k(t)))
\end{equation}
These equations have to be solved for a given trajectory $\vec R(t)$.
The square of the expansion amplitudes $a_{ij}$ taken at large
$t$ leads to the occupation probability for the state $j$.
In this way many experiments were described with high accuracy.
Figure 2 and 3 are typical examples and one could show dozens of
similar plots.
\begin{figure}[htb]
\epsffile{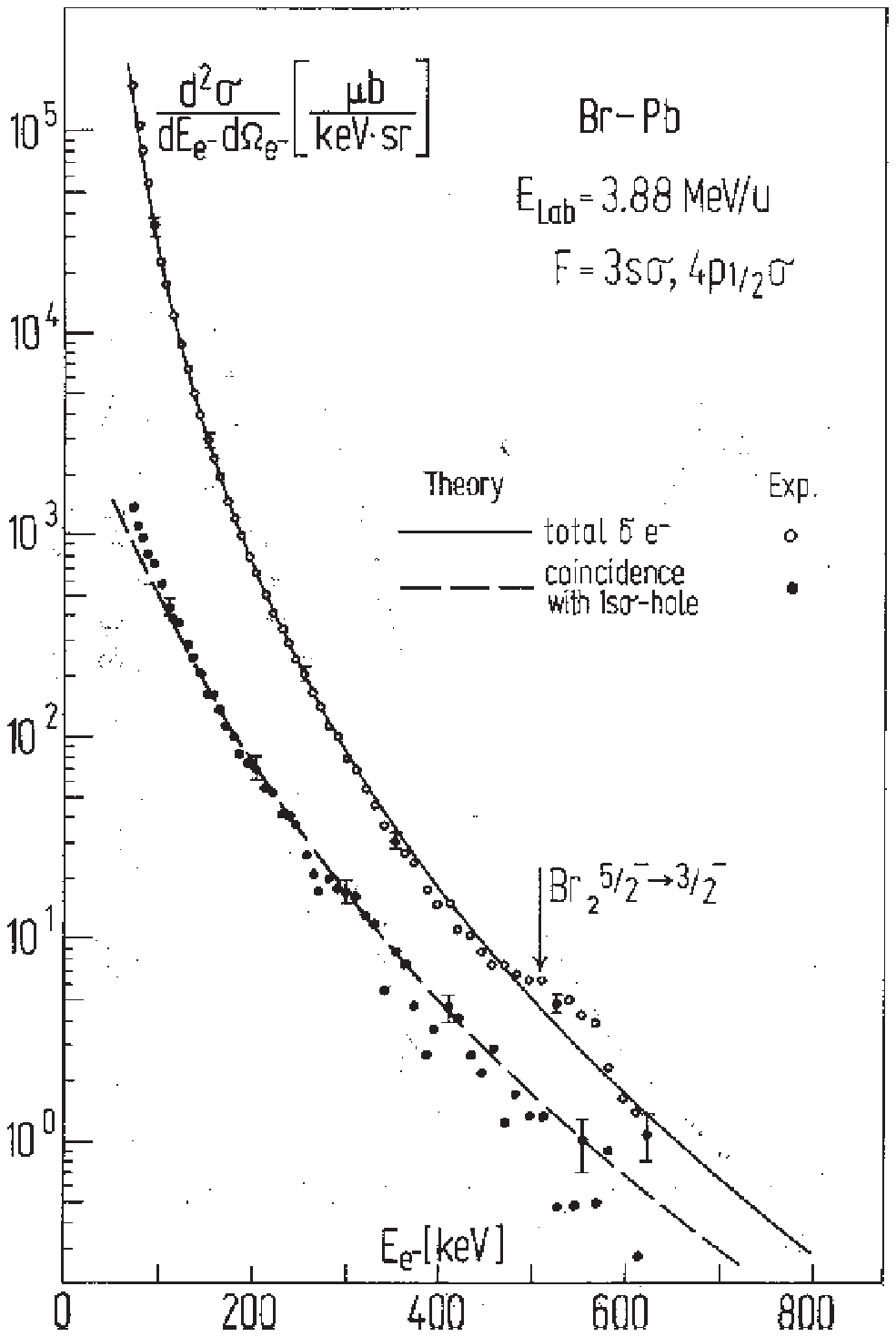}
\caption{$\delta$-electron spectra for Br on Pb. The agreement between
theory and experiment is perfect, up to the well-understood contribution
from an identified nuclear transition.}
\end{figure}
\begin{figure}[htb]
\epsffile{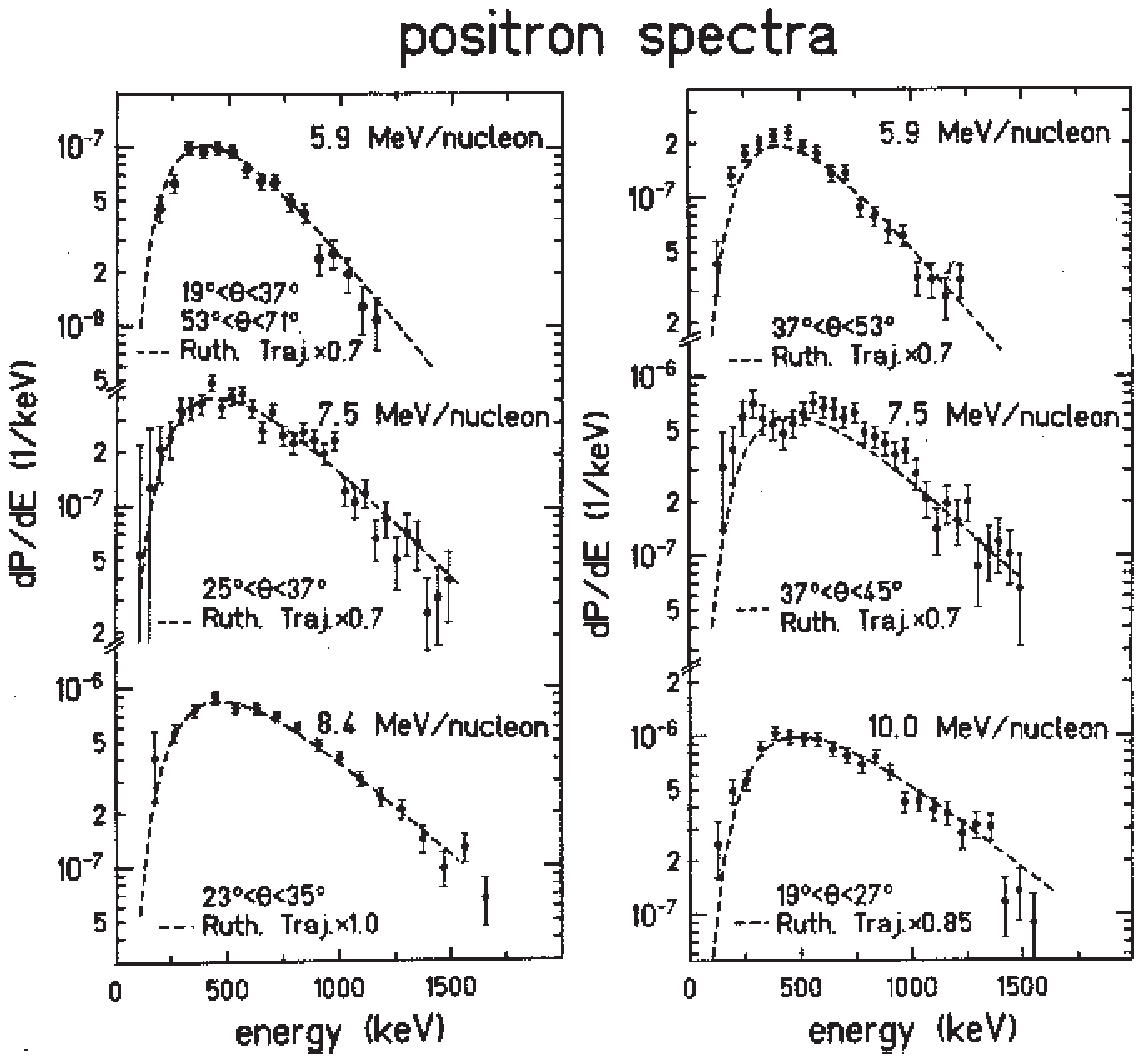}
\caption{Positron spectra for uranium on uranium for various
projectile energies. In each case a scaling factor between 0.7 and 1.0 was
introduced, which characterizes the absolut precision reached.}
\end{figure}
However, the most interesting process in Figure 1, namely the process c
which is called `spontaneous pair production' could not be observed so far
and theoretical calculations show that to observe it would require
the two ions to stick together much longer than usual.
It is generally agreed that for the rather broad class of nuclear reactions
summed over in the experiments so far, such long sticking times
cannot be expected. \\
In spite of this fact positron experiments at GSI found
pronounced positron lines \cite{exp1} and even extremely sharp
electron-positron
coincidence lines \cite{exp2}.
\begin{figure}[htb]
\epsffile{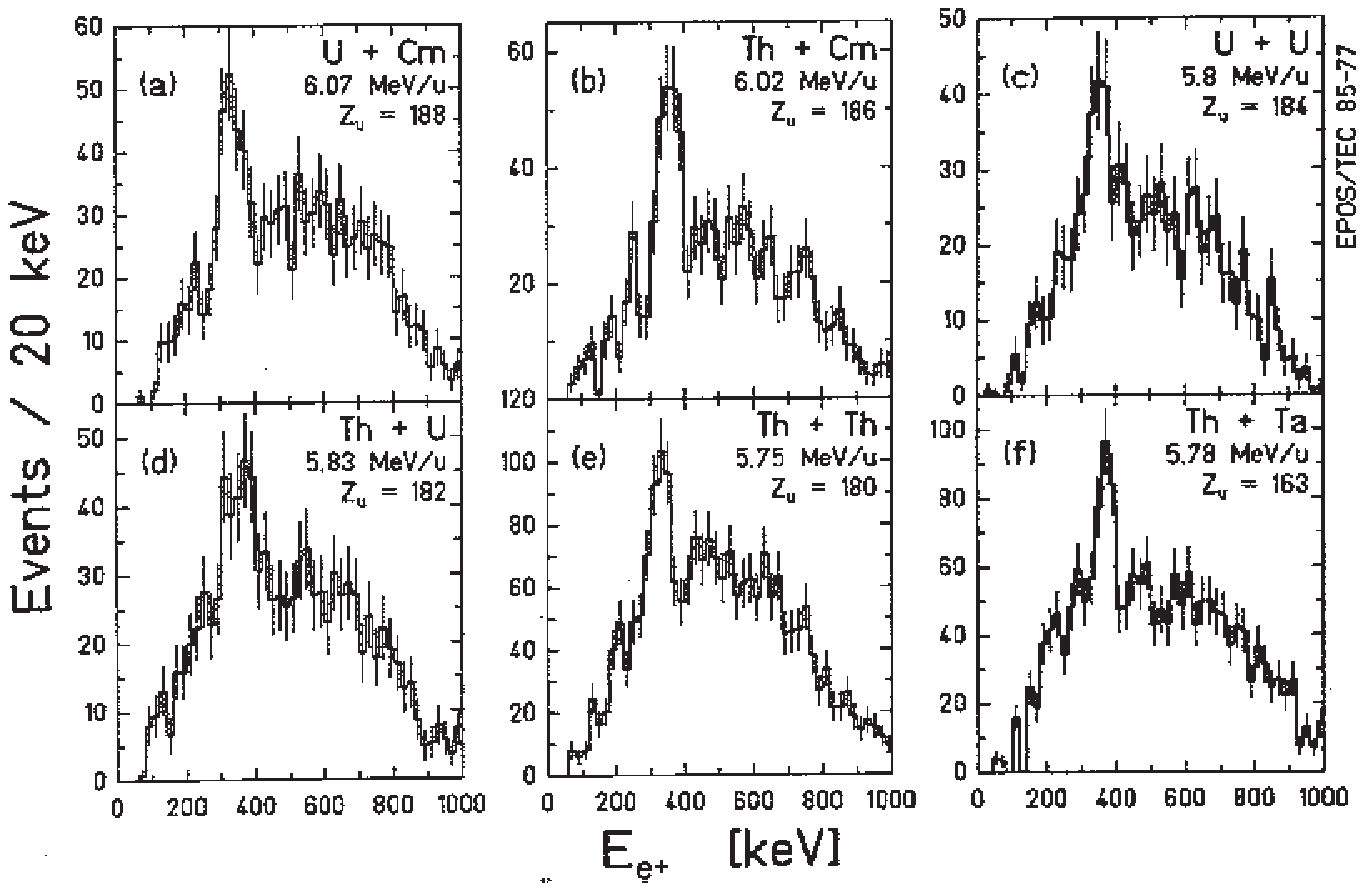}
\caption{The surprisingly large positron singles lines observed by EPOS in
1985}
\end{figure}
\begin{figure}[htb]
\epsffile{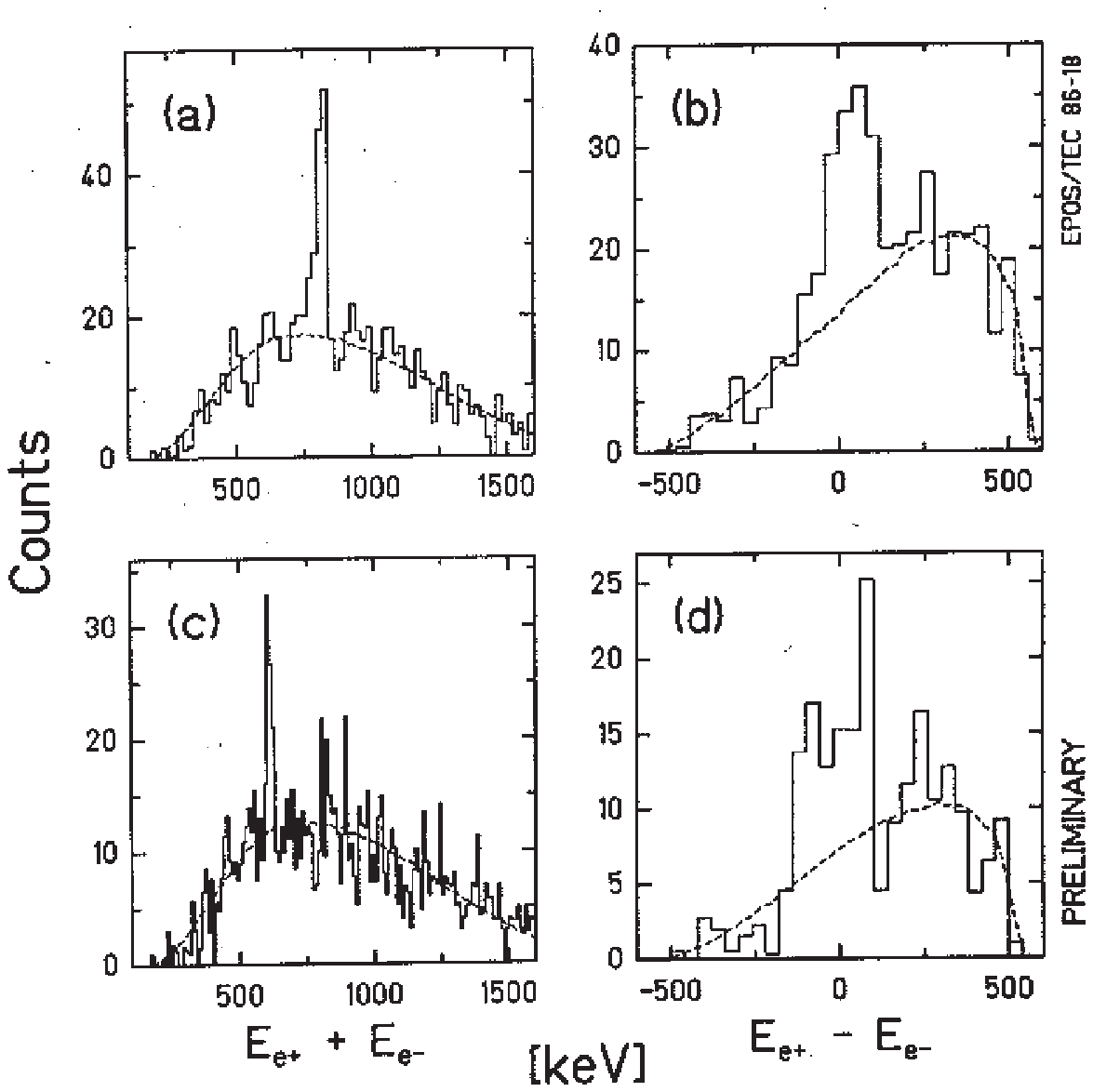}
\caption{The electron-positron coincidence lines observed by EPOS in 1988}
\end{figure}
At the time of their discovery it was tempting to interpret these
lines as signal for the decay of a new particle,
possibly the axion. New experiments and further theoretical investigations of
many low-energy phenomena showed, however, conclusively that this is no valid
possibility (for a review see \cite{as1}). We summarized these results already
early on as folllows:
`... we have to conclude that we did not find any scenario that would allow
to describe the production of an X particle with the required properties
without   very unnatural assumptions.' (1986, \cite{as2}) and
`... Therefore, it is tempting to conjecture that some of the experiments
are unreliable.' (1988, \cite{as3}).\\
The present situation  of these lines is the following:
Recent EPOS \cite{epos} and Orange \cite{orange}  experiments at
GSI could not reproduce the line observed
early although the detectors were improved to give much better statistics
(especially for EPOS). A completely independent experiment at Argonne National
Laboratory called APEX did not find any indication for lines \cite{apex}.
At the same time it became clear that nuclear pair-conversion was less well
understood than assumed, such that at least some of the data from recent years
might be due to conversion, a possibility which is still excluded for the
early
much larger line structure. It is unclear, how this has to be interpreted,
but definitely there is presently no compelling evidence for any process
more exotic than nuclear pair conversion.\\

\section{QED in strong fields}
A very interesting possibility opened by high-Z atoms is to test QED in a
non-perturbative regime. QED is a divergent field theory, and
the three basic divergences: vacuum polarization, self energy, and
vertex correction have to be renormalized. This renormalization procedure
in a  Coulomb-potential differs from that in free space, which results in
finite corrections to basically all physical quantities.
The complete understanding of these corrections is indispensible for a
reliable description of processes taking place in external fields.
In Frankfurt we
have a long tradition in calculating these \cite {Gr1}
and presently a number of projects are under work. Let me just
make a few comments on one of them, which is related to our work on
channeling.\\
{\em Delbr\"uck scattering}, i.e. the scattering of a photon off the
Coulomb-field
of an atom (see figure 6)
\begin{figure}[htb]
\epsffile{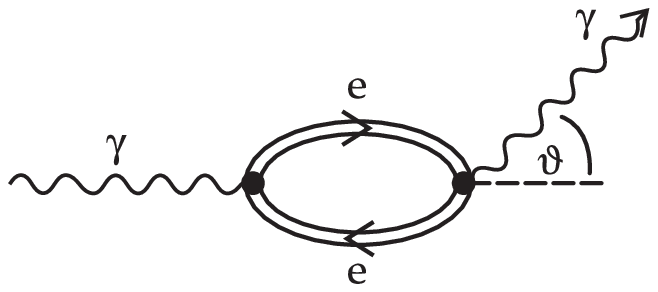}
\caption{The basic graph of Delbr\"uck-scattering. The double lines represent
exact propagators in the external field.}
\end{figure}
was carefully measured already some time ago \cite{Del1}.
Systematic deviations from the lowest-order theoretical prediction were found
for high-Z atoms, indicating that for these the higher order terms,
proportional to
$(Z\alpha)^{2n}$, are important. A direct perturbative calculation of the
next order graph is extremely involved and anyway one would like to calculate
this
graph exactly in all orders of $Z\alpha$. This is in fact possible if the
perturbative propagator is substituted by the exact propagator for the
Dirac-operator including the Coulomb-field
\begin{equation}
(i\gamma_{\mu}(i\partial^{\mu}-eV_{\rm Cb}(r)\delta_{\mu 0}-m)
S(x)=i\delta^4(x)
\end{equation}
We decomposed the propagator again into contributions of definite angular
momentum and inserted these into the calculation of the vacuum polarization
graph. The resulting calculation is very demanding, involving thousends
of angular momentum states, and requiring many numerical and analytic tricks
to ensure
convergence. Details can be found in ref. \cite{Del2}. Some results are given
in
the table  for photons with the energy $\omega=1.5 ~m_{\rm e}$
for the scattering angle $\Theta=1^{\circ}$
\begin{table}
\caption{Delbr\"uck scattering in all orders of $Z\alpha$}
\begin{tabular}{|c|c|c|c|c|}\hline
$\omega$ & $\Theta$ & $Z$ & $M_{\rm Born}$ & $M$\\
\hline
1.5~$M_{\rm e}$ & $1^{\circ}$ & 10 & 2.4$\cdot 10^{-3}$& $2.39 (3)\cdot
10^{-3}$ \\
1.5~$M_{\rm e}$ & $1^{\circ}$ & 20 & 9.6$\cdot 10^{-3}$& $9.62 (8)\cdot
10^{-3}$ \\
1.5~$M_{\rm e}$ & $1^{\circ}$ & 60 & 8.6$\cdot 10^{-2}$& $8.9 (1)\cdot
10^{-2} $\\
1.5~$M_{\rm e}$ & $1^{\circ}$ & 80 & 1.5$\cdot 10^{-1}$& $1.63 (2)\cdot
10^{-1}$ \\
\hline
\end{tabular}
\end{table}
Obviously one finds corrections with respect to the lowest-order result
for sufficently large $Z$. These investigations are still continued to
get complete systematic results for all photon energies and scattering
angles.\\

In another investigation we treated {\em channeling} by
solving the Dirac equation,
starting from first principles. Again this calculation was extremely involved.
However, it constitutes the first completely relativistic calculation for this
problem. If the physics of channeling were qualitatively similar to
that of strong fields this should show up in specific relativistic effects.
In fact we got rather satisfactory results (see figure 7 and 8,
and \cite{Kon,Me} ), but we found
\begin{figure}[htb]
\epsffile{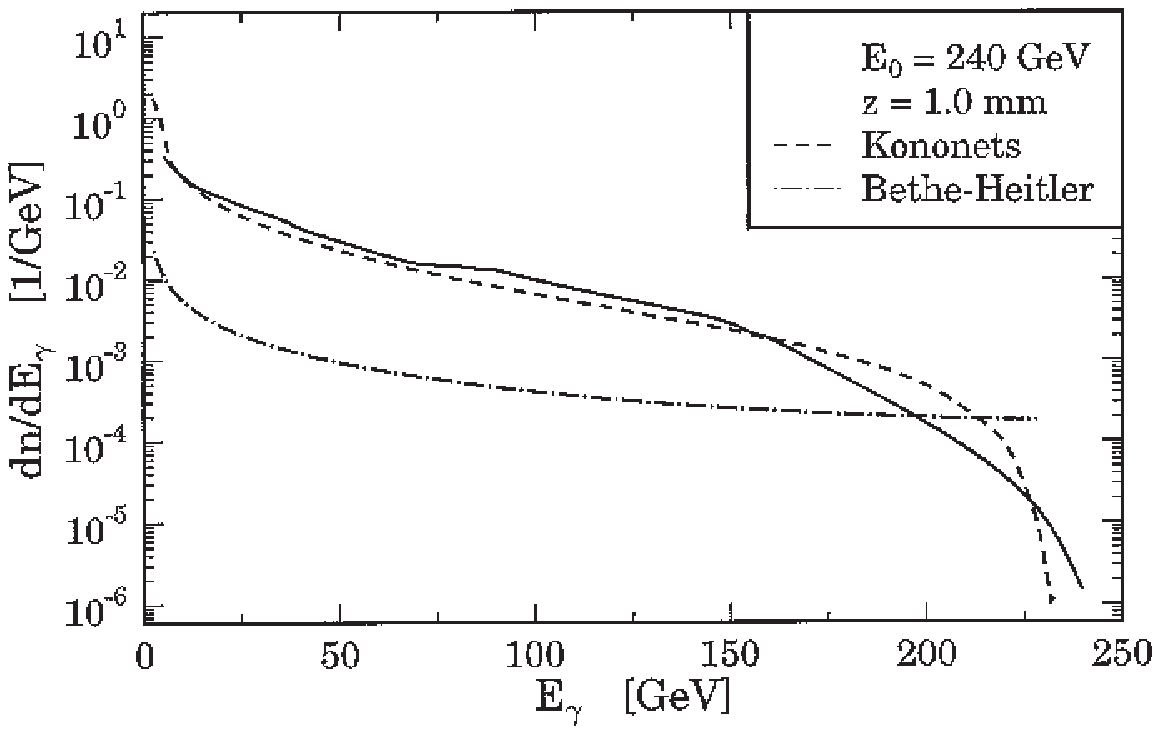}
\caption{Typical photon spectrum for relativisic channeling along the
$<110>$ axes
of Germanium, compared with the quasi-classical results of Kononets (ref. 12)
and the Bethe-Heitler formula. Obviously we confirm the Kononents results.
The lack of smoothness of our result is due to the finite number of states and
transition matrix elements taken into account.}
\end{figure}
\begin{figure}[htb]
\epsffile{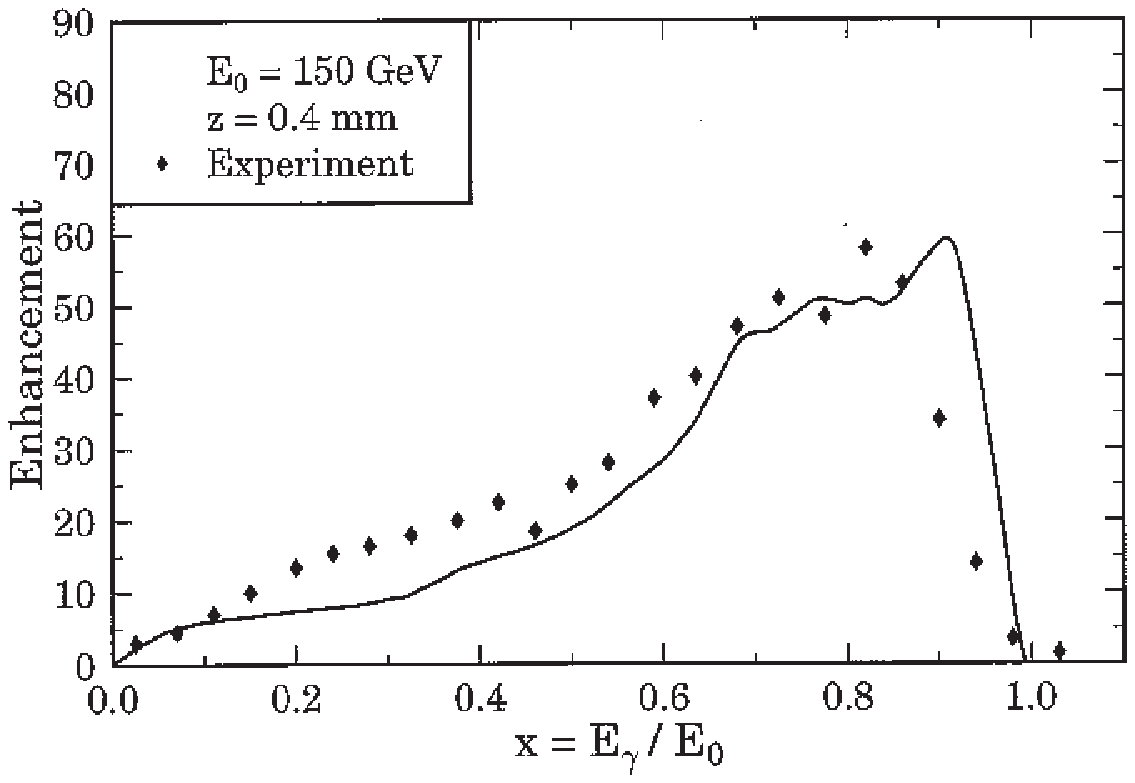}
\caption{Our result for the enhancement of fractional energy loss
during channeling in germanium, relative to an amorphous target,
compared with the data from ref. 13.}
\end{figure}
no specifically relativistic effects. It is usually argued that the
relevant mass is very large $m_{\perp}=\sqrt{p_z^2+m_{\rm e}^2}$, such that
one can perform
quasi-classical calculations. Our results confirm this and show at the same
time
that for  such effectively very heavy particles no phenomena occur which would
be similar to overcritical binding in heavy atoms.
As a by-product of this calculation we obtained the distribution of
the radial distance from the channeling axes at
which photons are emitted (see figure 9).
Again details can be found in our publication \cite{auge}.
\\
\begin{figure}[htb]
\epsffile{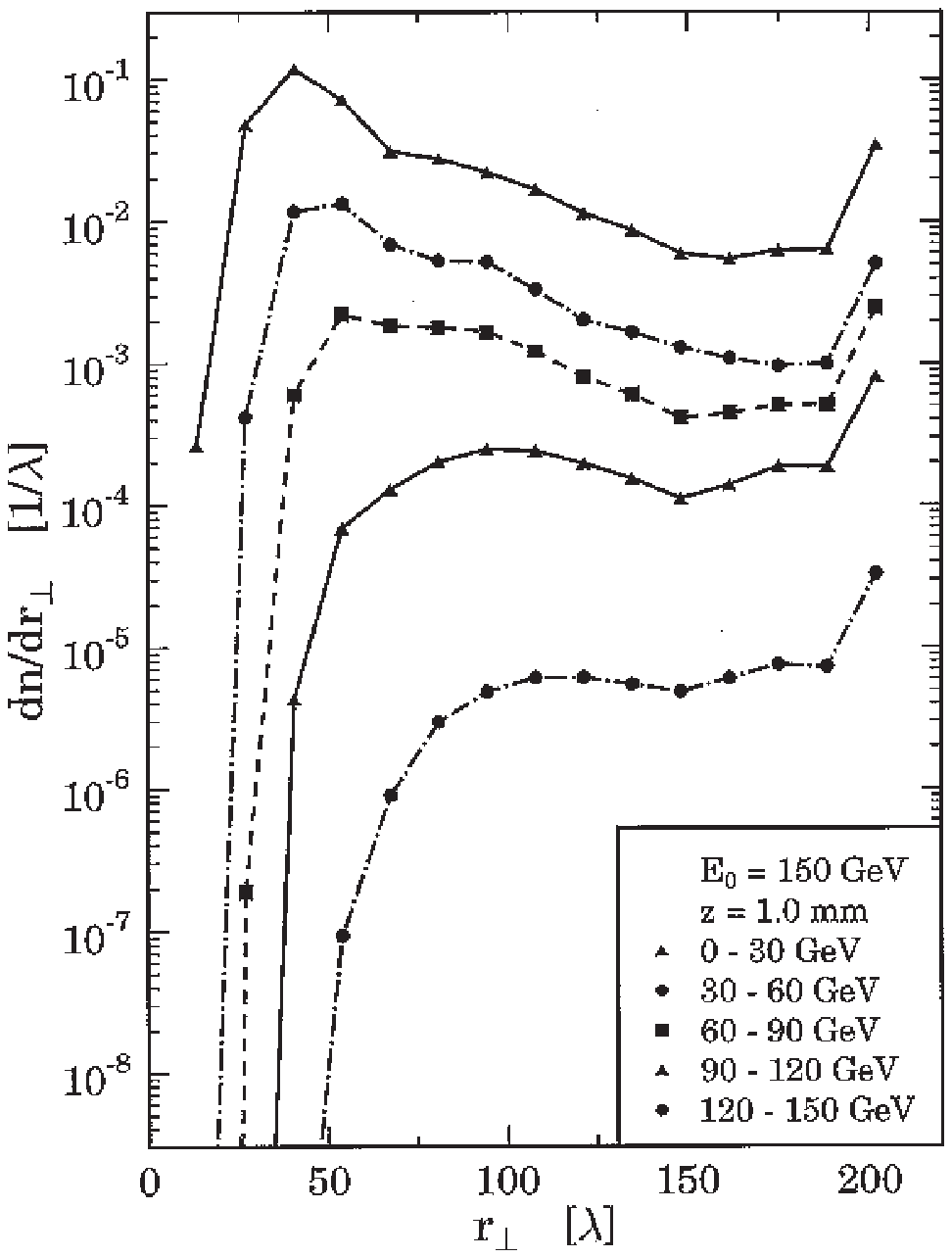}
\caption{Distribution of the emission points of photons as a function of the
radial distance from the channeling axis.}
\end{figure}
A similar conclusion was also reached by the following investigation.
We studied again the Delbr\"uck graph from figure 6, but this time for the
axially symmetric field of a channeling axis. The calculation was much
simplified
by the fact that this problem is two-dimensional rather than
three-dimensional.
The physical motivation was that a photon couples also to the
channeling fields due to
virtual production of electron-positron pairs, and this interaction could
lead to a
channeling of photons with a very small angle relative to the channeling axis,
see \cite{Klenner}.
One result is shown in figure 10, where the average rotation angle around the
\begin{figure}[htb]
\epsffile{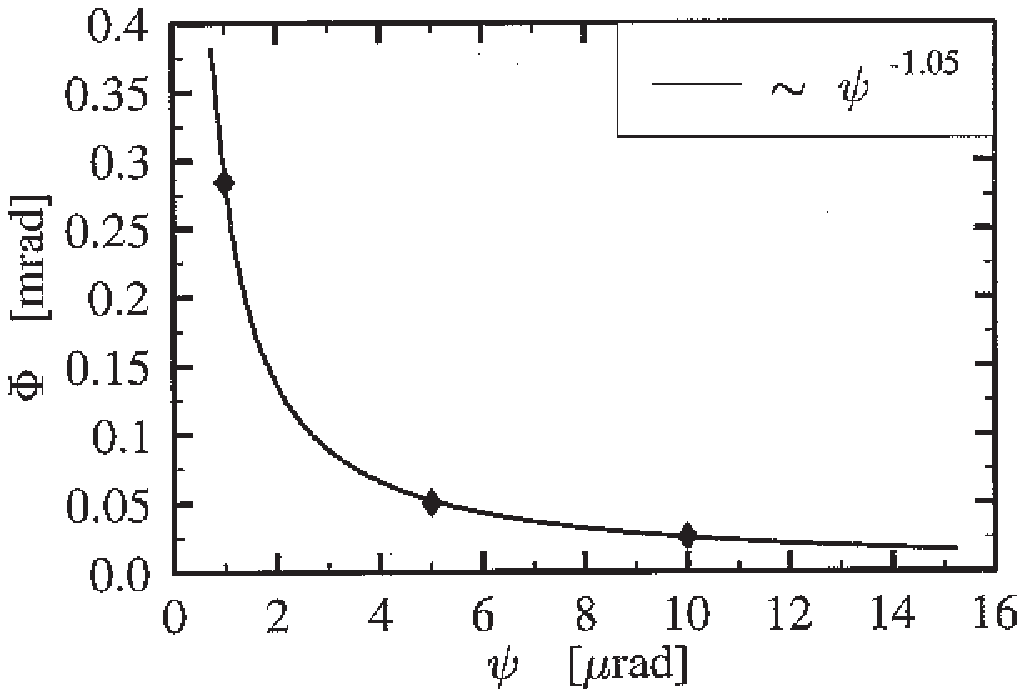}
\caption{The averaged angle by which a photon is scattered while passing
through a channel as function of the angle of incidence relative
to its axis.}
\end{figure}
channeling axis during the passage of a photon through the central part of the
channeling potential is shown as a function of the angle of incidence $\psi$.
$\Phi =\pi$ would imply complete channeling, i.e. the photon would be bound
to the channel. Obviously we are very far from this point, even for very small
angles. For details see \cite{Klenner}. Again, no substantial increase of
the effect with increasing photon energy was observed. Such an increase
would have indicated a prolonged existence of the
virtual electron-positron pair and thus a stronger
interaction with
the channeling potential) was observed.\\
Cutting a long discussion short, we
would like to argue that the channeling effect is basically  different
from the effects in over-critical Coulomb-fields for the following reason:
The fundamental process underlying all effects is the separation of an
electron-positron pair in a strongly space-dependent potential. The energy
gained by this charge separation is a measure for the size of the novel effets
up to the point where it allows the electron and positron to become real,
which would be spontaneous pair production. In the channeling situation the
electric fields becomes, however, only strong due to a Lorentz boost (into
the rest system of, e.g., the electron). The same boost introduces also an
magnetic
field which confines the electron and positron to a radial distance of the
size of a Landau orbit
$\Delta r\sim 1/B$. Thus the crucial quantity (electric field strength)
$\times$ (radial separation of electron and positron) is proportional to
$ E/B$, which is
basically
independent of the electron velocity, i.e. the Lorentz factor
$\gamma$. Thus no typical
overcritical field effects can be expected.\\
\begin{figure}[htb]
\epsffile{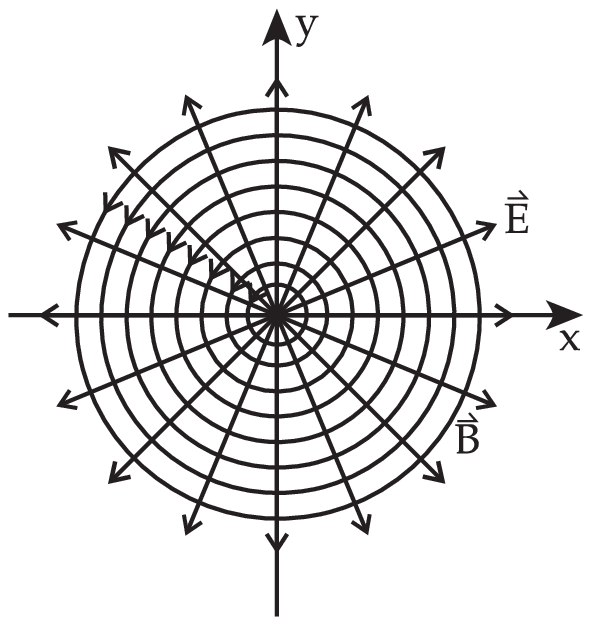}
\caption{The boosted electric and magnetic fields seen by a channeled
particle.}
\end{figure}

{\bf Acknowledgement:} I have reported results of the whole Frankfurt
atomic physics group, most notably J. Augustin, J. Klenner, U.
M\"uller-Nehler,
J. Reinhardt, A. Scherdin, G. Soff. This group is headed by  W. Greiner.
I also thank the experimentalists of the APEX, EPOS, and Orange collaboration
for communicating their results prior to publication.

\end{document}